\documentclass[12pt]{article}
 \usepackage{cite,eepic,epsf,epsf,epsfig,rotate,
multirow,verbatim,graphicx}
\newcommand{\newc}{\newcommand}
\newc{\tev}{\,{\rm TeV}}
\newc{\gev}{\,{\rm GeV}}
\newc{\sgn}{\mr{sgn}\,}
\newc{\ra}{\rightarrow}
\newc{\rpv}{$\mathrm{\not\!R_p}$}
\newc{\met}{$\not\!\!E_T$}
\newc{\rp}{$\mathrm{R_p}$}
\newc{\real}{\mathcal{R}e}
\newc{\alsm}{{\displaystyle \sum_{\alpha=1,2}}}
\newc{\besm}{{\displaystyle \sum_{\beta=1,2}}}
\newc{\al}{\alpha}
\newc{\ga}{\gamma}
\newc{\de}{\delta}
\newc{\cw}{\cos\theta_w}
\newc{\ssw}{\sin^2\theta_w}
\newc{\ccw}{\cos^2\theta_w}
\newc{\cbe}{\cos\beta}
\newc{\sbe}{\sin\beta}
\newc{\sh}{\hat{s}}
\newc{\sa}{\sin\al}
\newc{\ca}{\cos\al}
\newc{\bv}{$\mathrm{\not\!B}$}
\newc{\lv}{$\mathrm{\not\!L}$}
\newc{\ie}{{\it i.e.\/}\ }
\newc{\lam}{\lambda}
\newc{\cht}{\tilde{\chi}}
\newc{\upt}{\tilde{u}}
\newc{\elt}{\tilde{\ell}}
\newc{\hgt}{\tilde{H}}
\newc{\nut}{\tilde{\nu}}
\newc{\dnt}{\tilde{d}}
\newc{\psb}{\bar{\psi}}
\newc{\rtt}{\sqrt{2}}
\newc{\mut}{\tilde{\mu}}
\newc{\mr}{\mathrm}
\newc{\bath}{\bar{\theta}}
\newc{\tht}{\theta}
\newc{\JC}{{\bf J}}
\newc{\lra}{\longrightarrow}
\newc{\eg}{{\it e.g.\,}}
\newc{\barr}{\begin{array}}
\newc{\earr}{\end{array}}
\newc{\bary}{\begin{eqnarray}}
\newc{\eary}{\end{eqnarray}}
\newc{\dis}{\displaystyle}
\newc{\beq}{\begin{equation}}
\newc{\eeq}{\end{equation}}
\newc{\me}{\mathcal{M}}
\newc{\dbm}{\partial_\mu}
\newc{\sgm}{\sigma_\mu}

\def\ra{\rightarrow}
\def\dis{\displaystyle}
\def\bentarrow{\:\raisebox{1.1ex}{\rlap{$\vert$}}\!\rightarrow}

\def\dk#1#2#3{
        \begin{equation}
        \begin{array}{r c l}
        #1 & \rightarrow & #2 \\
         & & \bentarrow #3
        \end{array}
        \end{equation}
                }

\def\bothdk#1#2#3#4#5{
        \begin{equation}
        \begin{array}{r c l}
        #1 & \rightarrow & #2#3 \\
         & & \:\raisebox{1.3ex}{\rlap{$\vert$}}\raisebox{-0.5ex}{$\vert$}%
        \phantom{#2}\!\bentarrow #4 \\
         & & \bentarrow #5
        \end{array}
        \end{equation}
                }
\catcode`@=11 % This allows us to modify PLAIN macros.
\def \gsim{\mathrel{\mathpalette\@versim>}}
\def \lsim{\mathrel{\mathpalette\@versim<}}
\def \@versim#1#2{\lower0.4ex\vbox{\baselineskip\z@skip\lineskip\z@skip
     \lineskiplimit\z@\ialign{$\m@th#1\hfil##\hfil$%
     \crcr#2\crcr\sim\crcr}}}
\catcode`@=12 % at signs are no longer letters
\def\gev{\: \rm GeV}
\begin{document}
\setcounter{page}{0}
\renewcommand{\thefootnote}{\fnsymbol{footnote}}
\thispagestyle{empty}

\begin{titlepage}
%\vspace{-2cm}
\begin{flushright}
%IACS-P--000000\\[2ex]
%{\large \tt hep-ph/yymmnnn}\\
\end{flushright}
\vspace{+2cm}

\begin{center}
{ \Large \bf QCD corrections to excited lepton (pair) production at the LHC}\\
\vspace{+1cm}

\bf{Swapan Majhi}\footnote{E-mails:  tpskm@iacs.res.in}\footnote{Work supported
by CSIR Pool Scheme (Pool No. 8545-A)}\\

\it{Department of Theoretical Physics, \\
Indian Association for the Cultivation of Science \\
Kolkata 700032 India.}
\end{center}
\setcounter{footnote}{0}
\begin{abstract}\noindent
We consider 
the production of excited leptons 
($\bar{l^*}l$ as well as $\bar{l^*}l^*$) at the LHC,
 followed by their two body decay 
into Standard Model (SM) particles.
We perform the next-to-leading order (NLO)
QCD corrections to these processes. 
In spite of the non-renormalizable nature of the interaction,
such calculations
are possible and meaningful.
Not only are these corrections 
substantial and significant, 
the scale dependence of the NLO cross section is 
greatly reduced as 
compared to that for the
leading order (LO) cross sections.
\end{abstract}
\end{titlepage}

\setcounter{footnote}{0}
\renewcommand{\thefootnote}{\arabic{footnote}}

%%%%%%%%%%%%%%%%%%%%%%%%%%%%%%%%%%%%%%%%%%%%%%%%%%%%%%%%%%%
\section{Introduction}

The recent discovery of a Higgs like scalar particle at LHC may 
complete the most
successful model in particle physics namely the Standard Model (SM). 
In spite of this huge success, 
there are other issues like the replication of the fermion families, 
dark matter, baryogenesis etc. that are still not
understood within the framework of the SM. 
To address these, one needs to consider
physics beyond the SM. 
Some possible candidates are 
supersymmetry~\cite{susy}, grand
unification~\cite{Pati:1974yy,GUTS} (with or without supersymmetry),
family symmetries (gauged or otherwise) and  quark-lepton 
compositeness\cite{llmodel}.
The proliferation of fermion generations suggests the possibility of 
quarks and leptons being composite objects rather than elementary particles.
In these theories\cite{comp_mod, Additional_comp}, 
the fundamental constituents, {\em preon}s\cite{preon}, 
experience an additional strong and confining force. 
At energies far above a certain (compositeness) scale $\Lambda$,
preons are almost free.
Below this scale 
the interaction of preons become very strong forcing them to form 
bound states, namely quarks and leptons.
Understandably, in such models, higher (excited) states
          of quarks ($q^*$) and leptons ($l^* $) must also exist.
At energies below $\Lambda$, the interaction of the $l^*$ 
with the
SM fermions can
 be parametrized in terms of an effective four-fermion Lagrangian given 
by\cite{LagrangianCI} 
\beq 
{\cal L}_{CI} = {2 \pi \over \Lambda^2} {\displaystyle \sum_{i,j=L,R}}
\Bigg[
\eta_{ij} \Big(\bar{q}_i\gamma_{\mu}q_j\Big) \Big(\bar{l}_i^{*}\gamma^{\mu}{l}_j\Big) + 
\eta'_{ij}\Big(\bar{q}_i\gamma_{\mu}q_j\Big) \Big(\bar{l}_i^{*}\gamma^{\mu}{l}_j^{*}\Big) + 
h.c.\Bigg] \ ,
     \label{lagrangianC} 
\eeq 
where $l$ represents the SM lepton. 
In the above, we have not
explicitly accounted for the full $SU(2)_L \times U(1)_Y$ invariance
of the couplings, but this is to be understood, for the scale of
compositeness has to be larger than the electroweak scale. This implies 
that not only would we produce, say, $\bar e^{*} e^*$ and $\bar e^{*} e$, 
but also  $\bar e^{*} \nu_e^*$, $\bar e^{*} \nu_e$, and $\bar \nu_e^{*} e$.

 The excited fermions can also be transformed into ordinary SM fermions
through the gauge bosons. The effective gauge mediated 
Lagrangian\cite{LagrangianCI,LagrangianGM} 
between 
a SM fermion $F$ and its excited counterpart $F^*$
is given by
\beq
{\cal L}_{GM} = {1 \over 2 \Lambda} \bar{F^*_R} \sigma^{\mu\nu}
\Big[g_s f_s {\lambda^a \over 2} G^a_{\mu\nu}
+g f'' {\tau \over 2}.W_{\mu\nu} + g' f'{Y\over 2} B_{\mu\nu}
\Big]F_L + h.c.
     \label{lagrangianGM} 
\eeq
where $G_{\mu\nu}^a, W_{\mu\nu}$ and $B_{\mu\nu}$ are the field strength tensor
of the $SU(3)$, the $SU(2)$ and the $U(1)$ gauge fields respectively. 
The parameters $f_s,f''$ and $f'$ 
are usually of the order of unity.

It is evident that these operators may lead to 
significant phenomenological effects in collider experiments 
such as $e^+ e^-$~\cite{Delphi}, $e \, P$~\cite{HERA2} or 
hadronic\cite{cdfprl,CMS_CI,ATLAS_CI}. 
It is quite obvious that
the effects would be more pronounced at higher energies,  given the
higher-dimensional nature of ${\cal L}_{CI}$ and ${\cal L}_{GM}$. 
The best low-energy bounds on such composite operator would arise
from the precise measurement of leptonic branching ratios (BR)
of the $\tau$\cite{TauBR}. 
Similarly, loops with these excited
states can 
significantly modify rare processes and a comparison 
with the experimental data can impose bounds on 
their masses 
and couplings. 
These bounds, 
though, are quite weak \cite{TauBR_bounds}.
The 
best direct constraints on such excited states 
come from the Delphi~\cite{Delphi} and 
CDF~\cite{cdfprl} experiments. 
For the contact interaction scale $\Lambda = 1$ TeV, CDF
has excluded the excited electron mass below 756 GeV
at the $95\%$ C.L..
More recently, the 
measurement of the $\bar{l} l \gamma$ cross section\cite{CMS_CI, ATLAS_CI} 
at high invariant masses 
set
the most stringent limits on contact interactions 
of the type given in eqn.(\ref{lagrangianGM}). 
For $\Lambda = M_*$, CMS has excluded excited 
electrons below 1070 GeV 
and excited muons below 1090 GeV at the $95\%$ C.L..
For higher values of contact interaction scale (viz. $\Lambda = 2$ TeV),
the excited lepton mass has been excluded below 760 GeV for electrons
and 780 GeV for muons.

It is a well known fact that QCD corrections can alter, quite
significantly, generic cross sections at hadron colliders.  Even for a
simple process like Drell-Yan\cite{Drell_Yan}, the leading order (LO)
approximation is a serious underestimation, forcing us to incorporate
at least the next-to-leading order (NLO) or, better, next-to-leading
log (NLL)\cite{NLL,Martin} results in Monte Carlo codes \cite{NLL} or
event generators such as JETRAD\cite{JETRAD} and HERWIG\cite{HERWIG}.
It is expected that such corrections would be important in the context
of other processes as well.  Recently, the above-mentioned contact
interactions have received much attention from both CMS\cite{CMS_CI}
and ATLAS\cite{ATLAS_CI} collaborations.  They have searched for heavy
excited leptons via the $\bar{l} l\gamma$ channel and put a bound on
its mass.  However, there exists no higher order calculations for this
process, and, consequently, all collider searches of contact
interaction have either been based on leading order calculations, or,
have assumed that the higher order corrections are exactly the same as
for the SM process.
In this article, we aim to rectify this unsatisfactory
state of affairs.  While it may seem that the NLO corrections to the
processes driven by such non-renormalizable interactions are
ill-defined, it is not quite true\cite{SM_CI,ravi_gravi1}. In
particular, if the interaction can be factorized into 
a current-current form, with colored fields 
appearing in only one current, then the NLO QCD corrections
affect only this current and can be computed 
 without any difficulties. For example,
Ref.\cite{SM_CI} dealt with contact interactions involving the SM
fermions alone.

The rest of the article is organized as follows. In Section \ref{NLO
  corrections}, we start by outlining the general methodology and
follow it up with the explicit calculation of the NLO corrections to
the differential distribution in the dilepton
($\bar{l^*}l,~\bar{l^*}l^*$) invariant mass.  In Section \ref{results} we
present our numerical results. And finally, we summarize in Section
\ref{conclusion}.

%%%%%%%%%%%%%%%%%%%%%%%%%%%%%%%%%%%%%%%%%%%%%%%%%%%%%%%%%%%
\section{NLO corrections}
\label{NLO corrections}
        We consider excited leptons production 
in the context of contact interaction as exemplified by 
eqns.(\ref{lagrangianC},\ref{lagrangianGM}) at LHC. 
The processes are 
 
\dk{P(p_1)+P(p_2)}{{l^{*}}^{\!\!\!\!\!\!^{^{(-)}}}(l_1)+ {\bar{l}}^{\!\!\!^{^{^{^{(\,\, )}}}}}(l_2) + X(p_X)}{l^{\!\!\!\!\!\!^{^{(-)}}}(l_3)+V(p_4)\;,}
\bothdk{P(p_1)+P(p_2)}{\bar{l^{*}}(l_1)}{\,\,\,l^{*}(l_2)+ X(p_X)}{l(l_4)+V(p_5)}{\bar{l}(l_3)+V'(p_4)\;,}

where $p_i(i=1,2)$ denote the momenta of the incoming hadrons and
$l_i$ those for the outgoing leptons.  
Similarly, the outgoing vector bosons
  $V, V' (= \gamma,Z,W)$ have momenta $p_{4,5}$ whereas the inclusive
  hadronic final state carries $p_X$.  
In the above mentioned
processes, we have considered only two body leptonic 
decays of the excited leptons\footnote{
Also possible are 
three body decays through the four-fermi interactions with their
own QCD corrections. We postpone a discussion of this issue 
to a later study\cite{CI_3BDK}
}. 
The hadronic cross section is defined
in terms of the partonic cross section convoluted with the appropriate
parton distribution functions $f_a^P(x)$ and is given by \beq
%\barr{rclcrcl}
2 S {d\sigma^{P_1 P_2} \over d Q^2 } = \sum_{ab= q,\bar{q},g} \int_0^1 dx_1\: 
\int_0^1 dx_2 \: f^{P_1}_a(x_1) \: f^{P_2}_b(x_2) \, 
\int_0^1 dz ~2\, \hat{s}\; {d\sigma^{a b } \over d Q^2} \, 
\delta(\tau - z x_1 x_2),
%\\[2ex]\nonumber
%&&\times \Big({1\over s_{34}-m^2_{*}}\Big)^2 \;\Gamma(l^{*}\rightarrow l+\gamma)
%\earr
\label{eq:hadr_cross1}
\eeq
where $x_i$  is the fraction of the initial state proton's momentum 
carried by the $i^{th}$ parton. 
For the sake of completeness,
\beq
\barr{rclcrclcrcl}
S &\equiv& (p_1+p_2)^2 
& \qquad &  
\hat{s} &\equiv& (k_1+k_2)^2  
 & \qquad &  
Q^2 &\equiv&  (l_1+l_2)^2 
\\[2ex]
\tau&\equiv& \dis {Q^2 \over S}
& \qquad &  
 z &\equiv& \dis {Q^2 \over \hat{s} } 
& \qquad &  
\tau &\equiv& z\,x_1\,x_2 .
\earr
\eeq
At a first glance, the non-renormalizable nature of the effective 
 Lagrangian threatens to come in the way of reliably calculating loops. 
However, the fact that it can resolved into a product of a hadronic current 
with a non-hadronic one allows us to factorize the QCD corrections. 
These affect only the hadronic current, with the latter being a dimension three operator. With the leptonic tensor being a mute spectator, the offending 
higher-dimensional nature of the effective Lagrangian never comes into play.

Of particular interest is the leptonic tensor with
two massive final state particles, namely 
\beq
{ L}^{jj^{\prime} \rightarrow \,l\, l^{\prime} } =
\int \prod_i^{2}\Bigg({d^nl_i \over (2\pi)^n} \: 2 \pi
\,\delta^{+}(l_i^2-m^2_i)\Bigg) 
(2\pi)^n\, \delta^{(n)}\Big(q - l_1 - l_2\Big)
|{\cal M}^{jj^{\prime}\rightarrow \,l^* l^{\prime}}|^2 \ ,
\eeq
which leads to
 \beq
 {L}_{jj^{\prime} \rightarrow l^*\, l^{\prime} } 
    = 
       \dis\Big( -g_{\mu \nu} + {q_{\mu} q_{\nu} \over Q^2} \Big) \, 
                 {L}_{l^* l^{\prime}}(Q^2)   \qquad  
                    (l^{\prime} = l,l^*), 
\eeq
 with
 \beq
\barr{rcl c rcl}
 {L}_{l^* l}(Q^2) &=& \dis {1\over 12} \, 
                        \Big(Q^2 - {m^2_{1}+m^2_2 \over 2} - {(m^2_{1}-m^2_2)^2\over 2 Q^2}\Big).
\earr
 \eeq

To calculate the $Q^2$ distribution of the excited lepton pair ($\bar l^* l^*$
 or $\bar l^* l$), one  needs
to calculate the hadronic tensor as well. 
For this part of our calculation, we have 
followed the procedure of Ref.\cite{SM_CI}. 
The physical hadronic cross section can be obtained
by convoluting the finite coefficient functions with appropriate 
parton distribution functions and hence the inclusive differential 
cross section is given by 
\beq
\barr{rcl}
\dis 2 S {d\sigma^{P_1 P_2} \over dQ^2 }(\tau, Q^2) &=& \dis
\sum_q \int_0^1 dx_1\: 
\int_0^1 dx_2 
\int_0^1 dz ~ \delta(\tau - z x_1 x_2) 
\dis {\cal F}^{VA} \; {\cal G}_{VA} 
\\[3ex]
{\cal G}_{VA} & \equiv & 
H_{q\bar{q}}(x_1,x_2,\mu_F^2)\Big\{\Delta^{(0),VA}_{q \bar{q}}(z,Q^2,\mu_F^2)
+ a_s \Delta^{(1),VA}_{q \bar{q}}(z,Q^2,\mu_F^2) \Big\} 
\\[2ex]
&+ &
\Big\{H_{q g}(x_1,x_2,\mu_F^2) + H_{gq}(x_1,x_2,\mu_F^2)\Big\} a_s
\Delta^{(1),VA}_{q g}(z,\mu_F^2),
\label{dsig:dm}
\earr
\eeq
where the renormalized parton flux $H_{ab}(x_1,x_2,\mu_F^2)$ and 
the finite coefficient functions $\Delta^{(i)}_{ab}$ are given 
in Refs.\cite{ravi_gravi,ravi_gravi1,SM_CI}.
The effective coupling ${\cal F}^{VA}$ 
contains information of all the couplings, propagators and the massive final state particles 
and is given by 

\beq
{\cal F}^{VA} = 
{|\eta|^2\beta\over 12}{Q^2\over \Lambda^4} \Bigg\{ 
1-{(m_1^2+m_2^2)\over 2 Q^2} -{(m_1^2-m_2^2)^2\over 2 Q^4}\Bigg\},
\eeq
\beq
\beta = \Bigg(1 + {m_1^4\over Q^4}+ {m_2^4\over Q^4} - 2{m_1^2\over Q^2}
- 2{m_2^2\over Q^2} - 2{m_1^2\over Q^2}{m_2^2\over Q^2}\Bigg)^{1\over 2}.
\eeq

%%%%%%%%%%%%%%%%%%%%%%%%%%%%%%%%%%%%%%%%%%%%%%%%%%%%%%%%%%%
\section{Results and Discussion}
\label{results}

In the previous section, we have calculated the differential distribution 
with respect to the 
invariant mass of the leptonic pair  
(either $\bar{l^*}l$  or $\bar{l^*}l^*$).
The total cross section 
is trivially obtained
by integrating over $Q^2$ 
namely
\beq
\dis \sigma^{P_1 P_2} (M^2_{*},S,\Lambda) = 
\int {d\sigma^{P_1 P_2} (\tau,Q^2)\over d Q^2} d Q^2 .
\label{eq:Lsl_cs}
\eeq

We present our 
numerical results for three different LHC
energies, namely $\sqrt{S}= 7,8,14$ TeV.
We start by making the simplest choice for the renormalization 
and the factorization scale, viz. $\mu_R^2=\mu_F^2= Q^2$, and postpone
a discussion on the dependence on $\mu_{R, F}$ until later.
Since the QCD correction
does not depend on the contact interaction scale $\Lambda$, for definiteness
we 
use a particular value 
namely $\Lambda = 6$ TeV, 
unless 
it is quoted to be otherwise.
Similarly, all the coupling constants $\eta_{ij}$ and the $f$'s 
are also held to unity.
While the main findings of this paper are essentially independent of these 
specific 
values for the parameters, 
the later have been chosen so as to facilitate 
a quick and easy comparison with the experimental analyses existing 
in the literature.
For the same reason, we
use the Cteq6Pdf\cite{CTEQ6} parton distribution 
functions (PDFs),  
unless specifically mentioned otherwise.
Thus, the leading order (LO) hadronic
cross section is obtained by convoluting the LO parton distribution function
(namely Cteq6l1) with the
LO partonic cross section and
for the NLO hadronic cross section, we 
convolute  
the NLO parton
distribution (namely Cteq6m) with 
the NLO partonic cross section.
The corresponding QCD scale is
$\Lambda_{QCD} = 0.226(0.165)$ GeV 
for NLO (LO) for $n_f=5$.
%---------figure 1----------------------------------
\begin{figure}[htb]
%\vspace{-2cm}
\centerline{%\hspace*{3.0em}
\epsfxsize=18cm\epsfysize=8cm
                     \epsfbox{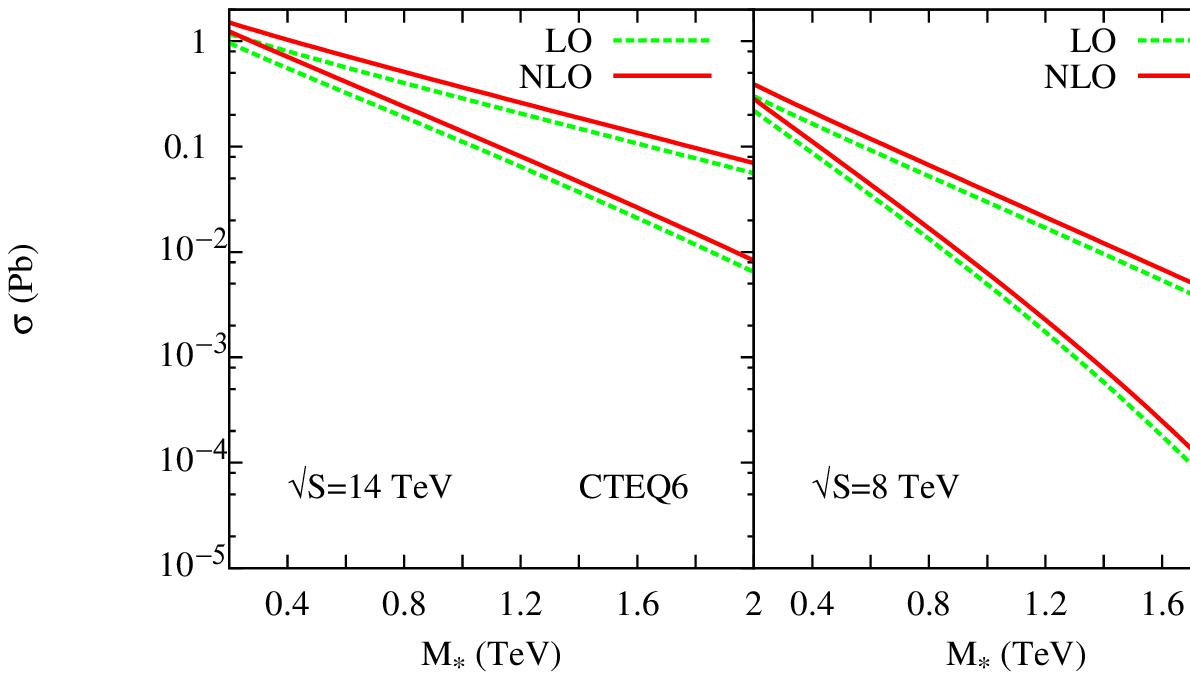}
}
\vspace{-1cm}
\caption{\em Variation of total cross-section for $\bar{l^*} l^* 
~{\rm and} ~\bar{l^*} l $ production with respect to 
excited lepton mass ($M_*$) at the LHC.
        For each set, the solid (dashed) lines refer to NLO (LO) cross
        sections. Upper (lower) set represents $\bar{l^*} l~(\bar{l^*} l^*)$ for $\Lambda = 6$ TeV.
        }
\label{fig:Ls_cs}
\end{figure}
%-------------

    To start with, we 
discuss the NLO corrections
to $\bar{l^*} l$ 
(this, by definition, includes $\bar{l} l^*$ as well)
and $\bar{l^*} l^*$ production in general, 
specializing later to a particular final state, namely $\bar{l} l \gamma$,
which
has been analyzed by both 
the CMS\cite{CMS_CI} and 
ATLAS\cite{ATLAS_CI} collaborations. 
In the context of the excited lepton, this 
final state is primarily attained through the production 
and subsequent decay of an $l^*$.
As the decay is free of any QCD correction,
the NLO QCD correction to the 
full process, namely
$\bar{l} l \gamma$ production 
is essentially the same as that for on-shell $\bar{l^*}$ production.
%----------------------figure 2------------
\begin{figure}[htb]
\vspace{-1cm}
\centerline{%\hspace*{3.0em}
\epsfxsize=18cm\epsfysize=12cm
                     \epsfbox{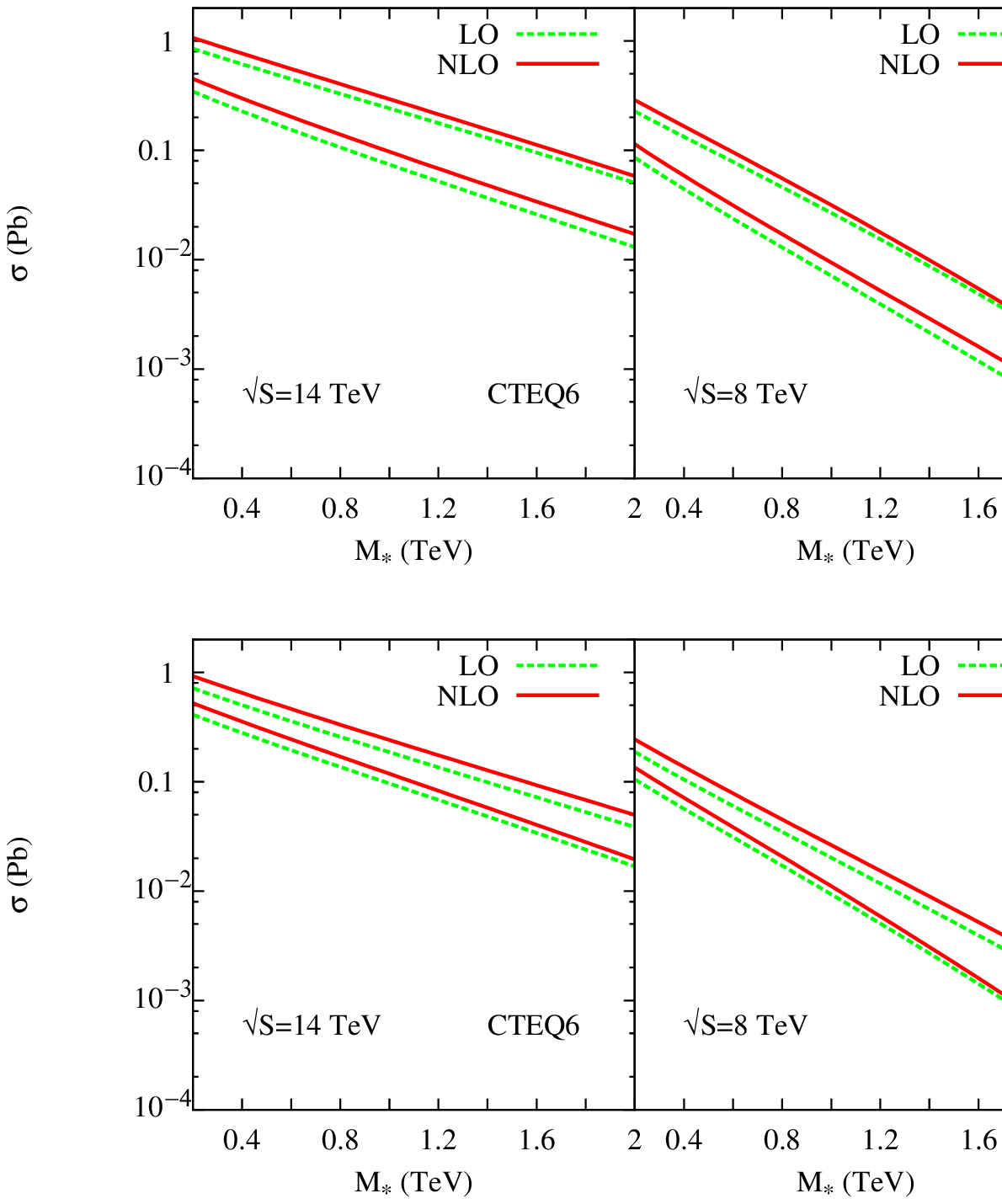}
}
\vspace{-1cm}
\caption{\em Variation of individual total cross-section for 
$\bar{l^*} l$ production with respect to 
the excited lepton mass ($M_*$) at the LHC.
        For each set, the solid (dashed) lines refer to NLO (LO) cross
        sections. In upper panel, the upper (lower) set represents 
$u\bar{u}~(d\bar{d})$ initiated process and
in lower panel, the upper (lower) set represents 
$u\bar{d}~(d\bar{u})$ initiated process at born level for $\Lambda = 6$ TeV.
        }
\label{fig:Ind_Ls_cs}
\end{figure}
%-------------

In figure \ref{fig:Ls_cs}, we have plotted the total cross section 
for both single and pair-production of excited leptons, 
as a function of its mass $M_* $. In calculating the same, we have 
assumed that the four-fermi operators are flavour-democratic, {\em i.e.},
the couplings $\eta_{ij}$ ($\eta'_{ij}$) are independent of the quark flavour. 
In other words, the cross sections in figure \ref{fig:Ls_cs} contain 
the contributions of all of the light quarks ($u, d, s$), with those 
of the heavier quarks being essentially negligible. 
The contribution of the individual light quark is depicted 
in figure \ref{fig:Ind_Ls_cs}.
The decrease of the cross sections with $M_*$ is not only due to 
the fall of the partonic cross sections, but also due to the 
fall in effective flux of the $q \bar q$ pair (relevant for both the 
LO and the NLO calculations) as well as the $q g$ pair (relevant for 
NLO alone) with increasing parton momentum fractions.
Understandably, the fall of the total cross section is
faster for lower center of mass (c.o.m.) 
energies $\sqrt{S}(\equiv 7,~ 8$ TeV) than the higher c.o.m. energy 
$\sqrt{S}(\equiv 14$ TeV).
As expected, the $\bar{l^*}l^*$ production cross section is 
both lower than and 
falls faster 
compared to the $\bar{l^*}l$ production cross section.
All the cross sections (figures \ref{fig:Ls_cs}, \ref{fig:Ind_Ls_cs})
have similar qualitative features (though the actual
numbers are quite different), a reflection of the flavour-independence 
of the underlying dynamics. 

%-------------------------------------------
\begin{figure}[htb]
\vspace{-0.5cm}
\centerline{%\hspace*{3.0em}
\epsfxsize=18cm\epsfysize=7cm
                     \epsfbox{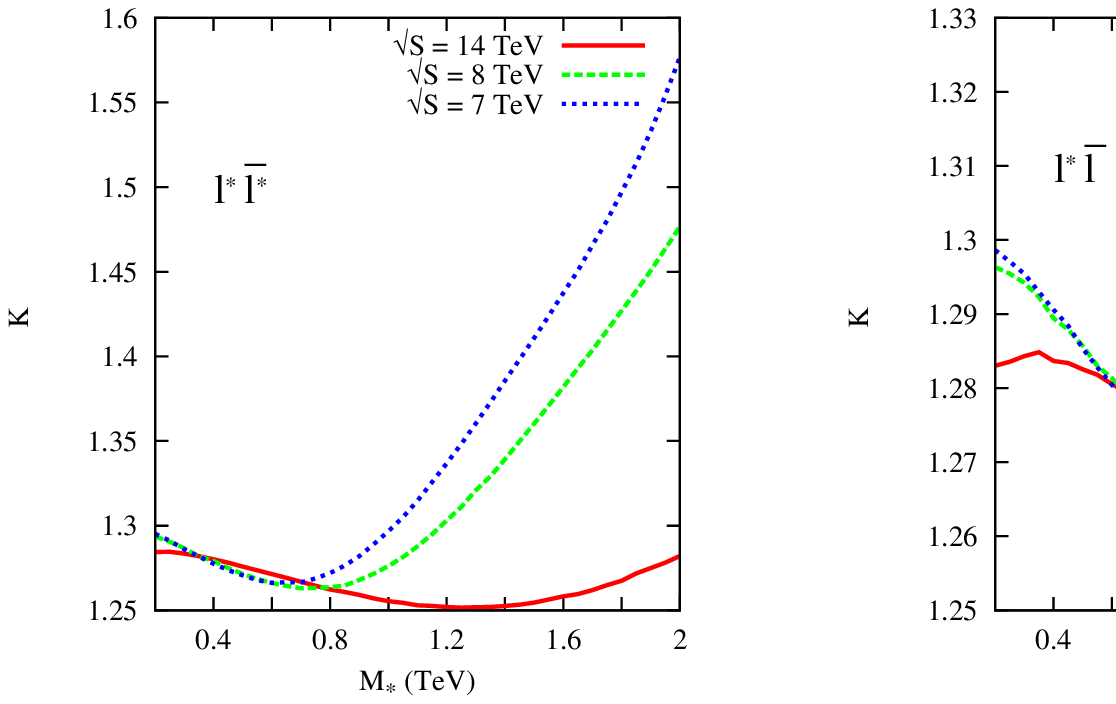}
}
\vspace{-0.5cm}
\caption{\em Variation of $K$-factor with respect to the excited lepton
mass ($M_*$) for $\Lambda = 6$ TeV at the LHC.
        }
\label{fig:KLs}
\end{figure}
%-------------
%-------------------------------------------
\begin{figure}[htb]
%\vspace{-2.0cm}
\centerline{%\hspace*{3.0em}
\epsfxsize=18cm\epsfysize=7cm
                     \epsfbox{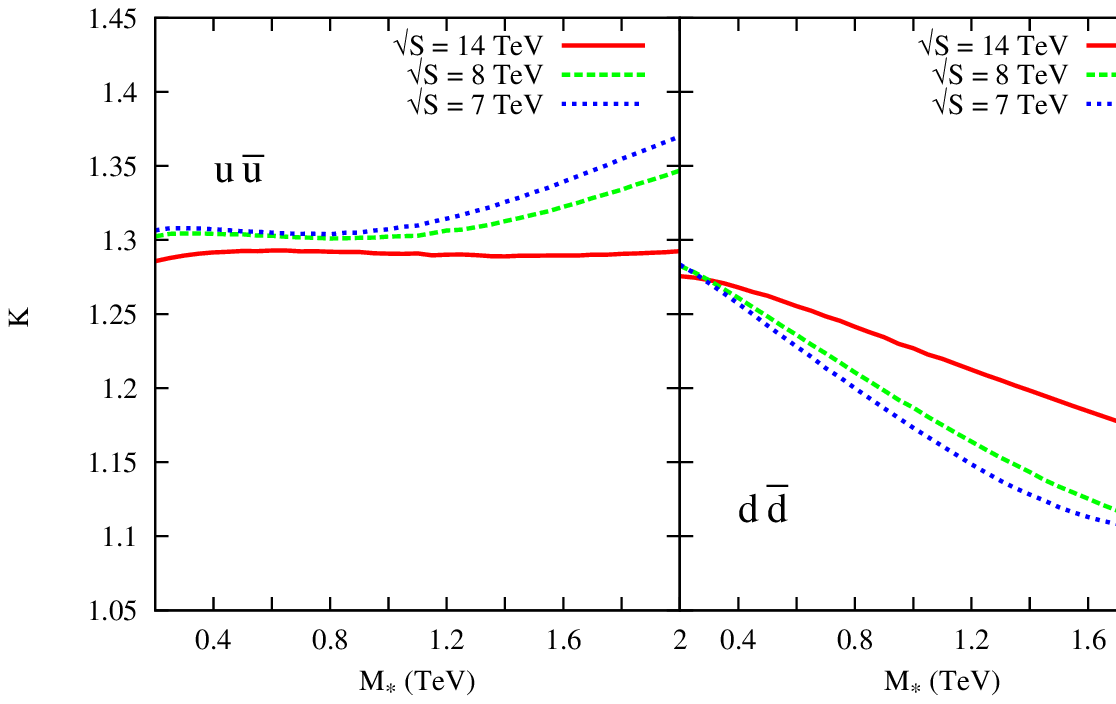}
}
\vspace{-0.5cm}
\caption{\em Variation of individual $K$-factor with respect to excited lepton 
mass ($M_*$) for $\Lambda = 6$ TeV at the LHC.
        }
\label{fig:KLs_Ind}
\end{figure}
%-------------

   To quantify the enhancement of NLO cross section, we define a variable
called $K$-factor as given by
\beq
K_i = {\sigma_{i}^{NLO} \over \sigma_{i}^{LO}}~~~ i = {\rm total}, u\bar{u},d\bar{d},
u\bar{d},d\bar{u}
\eeq
where the LO (NLO) cross sections are computed by convoluting the corresponding 
parton-level cross sections with the LO (NLO) parton distribution functions.

In figures (\ref{fig:KLs},\ref{fig:KLs_Ind}) we have 
shown the variation of 
the $K$-factor with respect to $M_*$.
The variation of the total $K$-factor is about $25\%-30\%$  
 for moderate values of $M_* (\leq 1$ TeV) at low c.o.m energies 
($\sqrt{S} = 7,~8$ TeV) in figure \ref{fig:KLs}. 
At larger mass region ($M_* > 1$ TeV), 
the $K$-factor rises very fast ($25\% - 60\%$). At high c.o.m energy
($\sqrt{S} = 14$ TeV), the variation of the $K$-factor is about $25\%-30\%$ 
for even larger masses ($M_* \leq 2$ TeV).
Figure \ref{fig:KLs_Ind} shows the variation of the $K$-factor 
for individual flavors only for the $\bar{l^*} l$ 
production process.
In figures (\ref{fig:KLs},\ref{fig:KLs_Ind}), the rate of change 
of the $K$-factor is much slower at higher
c.o.m energy (say $\sqrt{S} = 14$ TeV) than the lower c.o.m energies. 
This is a consequence of the fact that at lower c.o.m 
energies, we are forced to higher momentum fractions, and, hence,
 are integrating over smaller phase space regions. 
As the Bjorken $x$ increases towards unity, the parton distribution function
 falls very steeply
 This is the reason why at lower c.o.m. energies, 
the $K$-factor increases very fast as mass $M_*$ increases towards 
the center of mass energy.
One can also see from figure \ref{fig:KLs_Ind} that the  
numerical difference between the
individual flavor  $K$-factors is due to their respective flux difference.
Since the $d$-quark parton density falls faster than the $u$-quark parton density with scale,
the $K$-factor falls steeply for $d\bar{d}$ initiated processes than $u\bar{u}$
initiated processes. This also explains the variation of $K$-factor for $d\bar{u}$ and $u\bar{d}$
processes where the earlier processes, the flux dominated by valence $d$-quark 
and later, the flux dominated by valence $u$-quark. 

%####################
\subsection{$\bar{l}l\gamma$ production}
\label{llg}
%###################

The excited heavy lepton will decay into a light SM lepton and a electroweak 
gauge bosons $V(\equiv \gamma, Z, W)$ according 
to the Lagrangian of eqn.(\ref{lagrangianGM}).
Therefore the total NLO cross section of lepton pair ($\bar{l}l $) 
and a gauge boson $V$ 
can be calculated by multiplying the branching ratio to the 
eqn.(\ref{eq:Lsl_cs})  as given below 
\beq
\dis \sigma^{P_1 P_2} (M^2_{*},S,\Lambda) = BR(l^*\rightarrow l V) 
\int {d\sigma^{P_1 P_2} (\tau,Q^2)\over d Q^2} d Q^2 .
\eeq

The partial decay width of the excited lepton for various 
electroweak gauge bosons is given by
\beq
\Gamma(l^*\rightarrow l V) = {1\over 8} \alpha \,f^2_V {M_*^2 \over \Lambda^2} \Bigg(1-{m^2_V\over M^2_*}\Bigg) \Bigg(2+{m^2_V\over M^2_*}\Bigg),
\eeq
with
\bary
f_{\gamma} &=& f''\,T_3+f'{Y\over 2},\\
f_{Z} &=& f''\,T_3\cot\theta_W-f'{Y\over 2}\tan\theta_W,\\
f_{W} &=& {f''\over \sqrt{2}}\csc\theta_W, 
\eary
where $T_3$ denotes the third component of the weak isospin and $Y$ represents
the weak hypercharge of excited lepton. $\theta_W$ is the Weinberg's angle.
The compositeness parameters $f''$ and $f'$ are taken to be unity through
out our analysis. 
The variation of these parameters have been considered 
elsewhere (for example in Refs.\cite{PRD65,PRD81}).
The decays of excited lepton mediated by electroweak interaction is mostly
dominated by $W$-boson and a SM lepton.
For sufficiently large excited lepton mass
(at least larger than $m_W$ and $m_Z$), the branching ratios are insensitive to
$M_*$. However this is not quite true when one considered the three body decay 
through contact interactions. 
In this case, the decay width of contact interaction ($\Gamma_{CT}$) is 
dominated over the width of electroweak interaction ($\Gamma_{G}$) as 
the mass of excited lepton increases
which is shown in table \ref{tab:BrFrac}.
%%%%%%%%%%%%%%%%
\begin{table}[htb]
\begin{center}
\begin{tabular}{|c|c|c|c|c|}
\hline
 $M_*$ (GeV) & $\Gamma_{tot}/M_* $ & $\Gamma_G/\Gamma_{tot}$ & $\Gamma_{CT}/\Gamma_{tot}$ & Br($l^* \rightarrow l\gamma$)\\
\hline
 400 & 3.85$\times 10^{-4}$ &0.6557 & 0.3443 & 0.1894 \\
 600 & 1.25$\times 10^{-2}$ &0.4649 & 0.5351 & 0.1308 \\
 800 & 3.17$\times 10^{-2}$ &0.3303 & 0.6697 & 0.0911 \\
 1000 & 6.82$\times 10^{-2}$ &0.2407 & 0.7593 & 0.0668 \\
 2000 & 8.94$\times 10^{-2}$ &0.0738 & 0.9262 & 0.0204 \\
\hline
\end{tabular}
\caption[] {Decay widths of excited lepton and branching ratio $BR = \Gamma(l^* \rightarrow l \gamma)/\Gamma(l^* \rightarrow all)$
for $\eta_{ij} = f''=f'=1$ and $\Lambda = 2$ TeV. $\Gamma_{tot}$ represents the total decay width.}
\label{tab:BrFrac}
\end{center}
\end{table}
%%%%%%%%%%%%%%%
%-------------------------------------------
\begin{figure}[htb]
\centerline{%\hspace*{3.0em}
\epsfxsize=18cm\epsfysize=14cm
                     \epsfbox{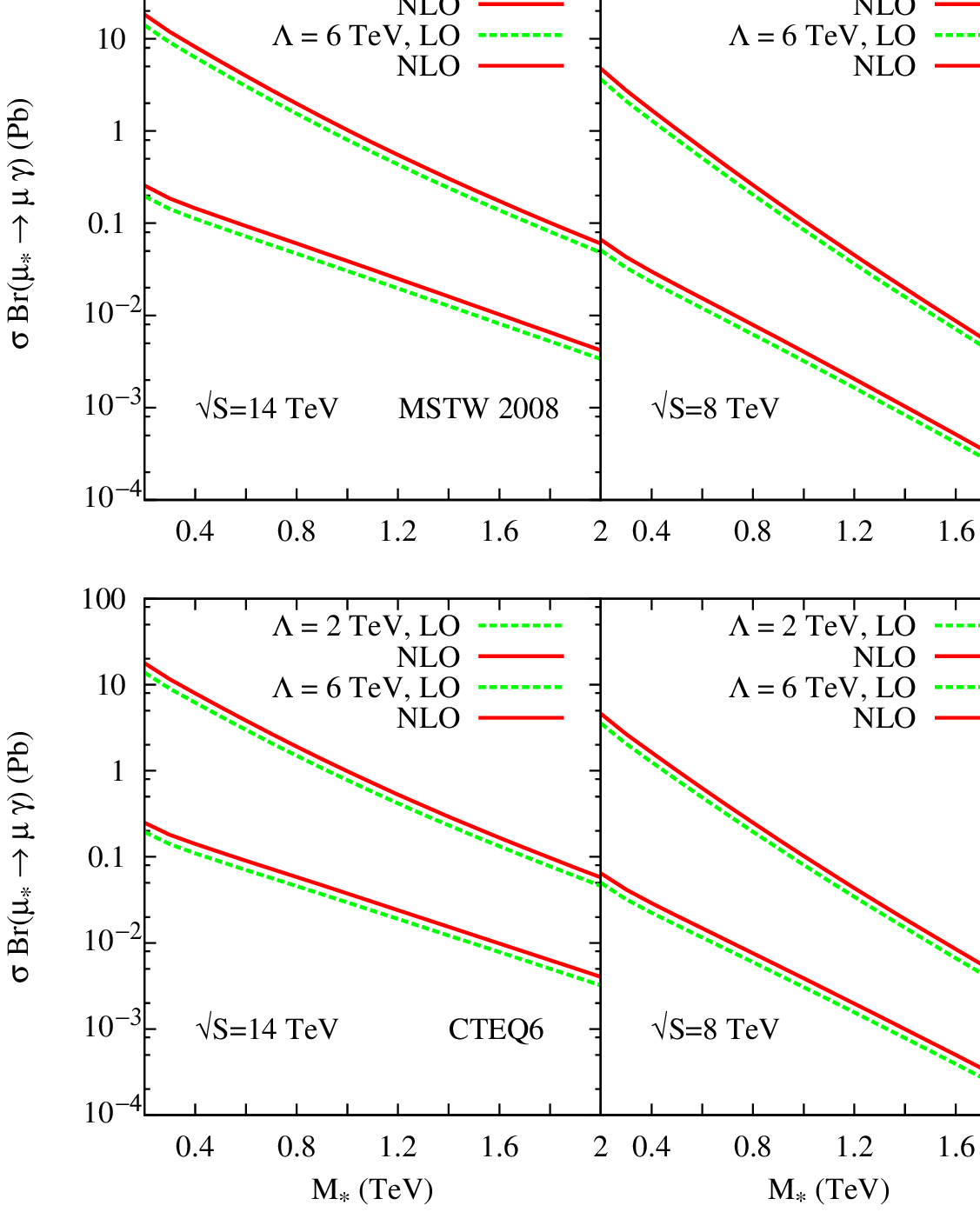}
}
\vspace{-1cm}
\caption{\em Total cross-section for $l \bar{l}\gamma$ production
             at the LHC.
        For each set, the solid (dashed) lines refer to NLO (LO) cross
        sections. Upper (lower) set is for $\Lambda = 2 (6)$ TeV.
        }
\label{fig:llg_cs}
\end{figure}
%-------------

In figure \ref{fig:llg_cs}, we have plotted the total cross section versus 
the invariant mass of a SM lepton and a photon 
$M_* (\equiv M_{l\gamma})$
for two different PDFs, namely CTEQ6\cite{CTEQ6} and MSTW 2008\cite{MSTW} 
for two different values of the contact 
interaction scale ($\Lambda = 2,~6$ TeV). 
As before (and for identical reasons), the cross section decreases as the
invariant mass $M_*$ increases. 
From figure \ref{fig:llg_cs},                              
we see that as the contact interaction scale ($\Lambda$) increases, the cross
section (both LO as well as NLO) decreases uniformly as $\Lambda^{-4}$ as 
expected from eqn.(\ref{lagrangianC}). 
Therefore, one can obtain the
 cross section (for both LO as well as NLO) for arbitrary values of 
$\Lambda$ by multiplying our results by 
an appropriate scale factor.

\begin{figure}[htb]
\vspace{-1.5cm}
\centerline{%\hspace*{3.0em}
\epsfxsize=18cm\epsfysize=12cm
                     \epsfbox{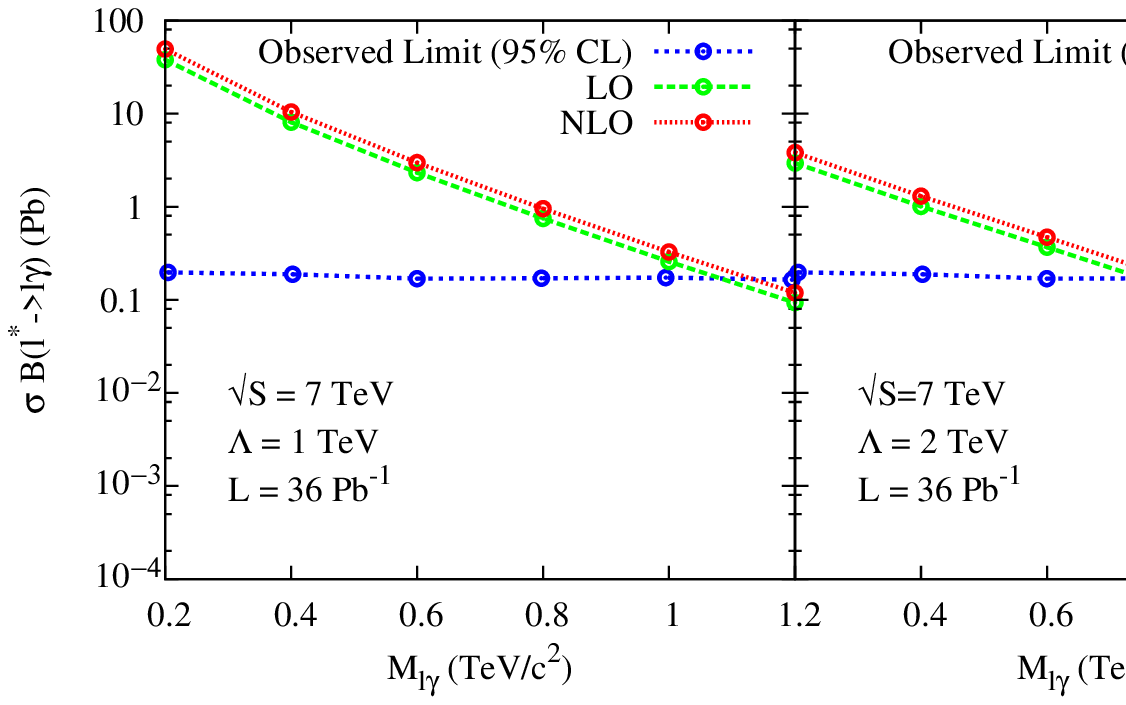}
}
\vspace{-1.5cm}
\caption{\em Variation of production cross section times branching fraction with
excited lepton mass (Red and Green lines) and the blue line is the observed 
limit (95\% CL) taken from CMS\cite{CMS_CI}.
        }
\label{fig:mass_bound}
\end{figure}

In Fig.\ref{fig:mass_bound}, we plot a particular measurable, viz. 
the product of the 
cross section and the branching 
fraction, along with the 95\% CL upper limit as obtained 
by 
the  CMS collaboration\cite{CMS_CI}.
From this figure, 
it is clear
 that on inclusion of NLO QCD corrections, the mass limit 
on the excited leptons
are enhanced 
somewhat, and which
we quantify 
in table \ref{tab:mass_limit}.  
%%%%%%%%%%%%%%%%
\begin{table}[htb]
\begin{center}
\begin{tabular}{|c|c|c c|}
\hline
$\Lambda$ (TeV) & $\sigma$ (Pb) & Excited lepton mass &(GeV)  \\
        &            &     LO    & NLO         \\
\hline
1       & 0.173 & 1077 (1070) & 1137 \\[1ex]
2       & 0.174 & 748 (760) & 804 \\
\hline
\end{tabular}
\caption[] {Mass limit for the excited lepton.
The number within the first bracket represents the CMS result.}
\label{tab:mass_limit}
\end{center}
\end{table}
%%%%%%%%%%%%%%%

\begin{figure}[htbp]
%\vspace{0.5cm}
\centerline{%\hspace*{3.0em}
\epsfxsize=18cm\epsfysize=22cm
                     \epsfbox{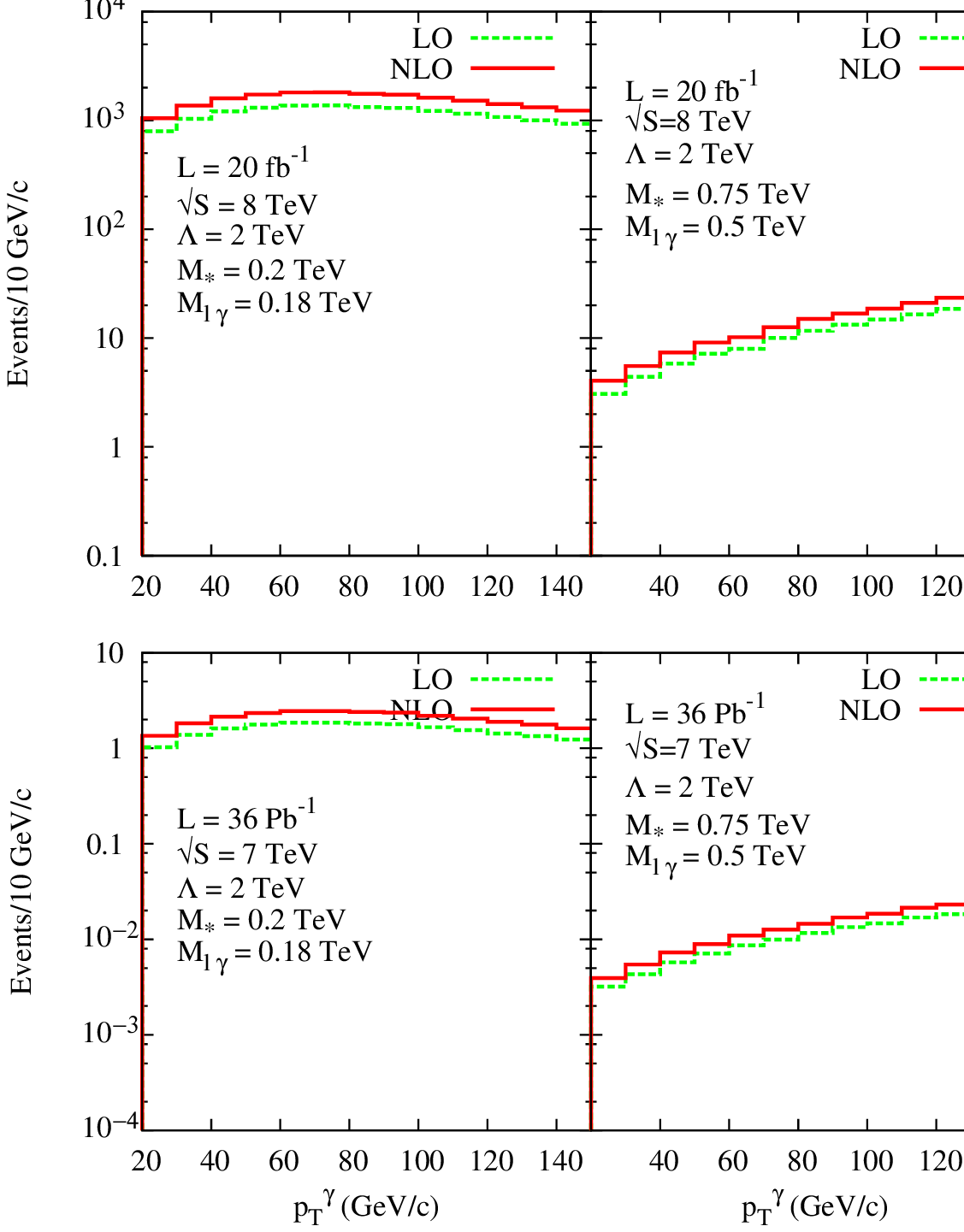}
}
\vspace{-1cm}
\caption{\em Photon transverse momentum distribution at three different
excited lepton masses and three different LHC energies for MSTW 2008 parton distribution functions.
        }
\label{fig:pT_dist}
\end{figure}

In figure \ref{fig:pT_dist}, we 
display the photon transverse momentum distribution 
with same lepton-photon invariant mass cut ($M_{l\gamma}^{cut}$) as given
 in \cite{CMS_CI}. 
In this figure, 
we consider the projected luminosity 
20 (100) $fb^{-1}$ at $\sqrt{S} = 8 (14)$ TeV LHC energy.
This figure demonstrates that the enhancement has a 
relatively small dependence on the photon $p_T$, and thus the 
language of the $K$-factor is a useful one not only for 
effecting Monte-Carlo studies of the process, but also for 
analysing actual data.

%%%%%%%%%%%%%%%%%%
%%%%%%%%%%%%%%%%
%%%%%%%%%%%%%%%
%%%%%%%%%%%%%%%

%%%%%%%%%%%%%%%
%-------------
\begin{figure}%[htb]
%\vspace{-2cm}
\centerline{%\hspace*{3.0em}
\epsfxsize=17cm\epsfysize=8.0cm
                     \epsfbox{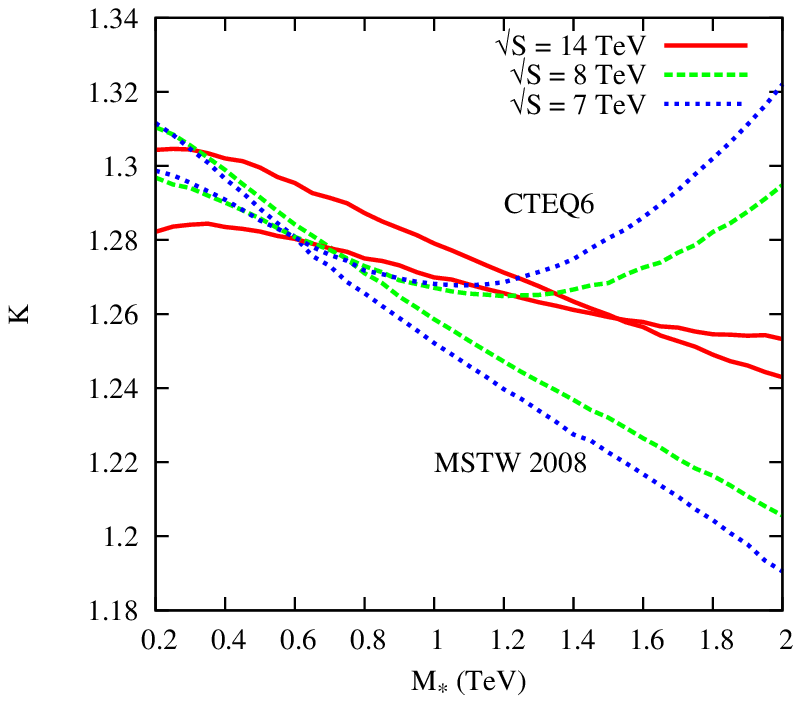}
}
\vspace{-0.5cm}
\caption{\em $K$-factor for $l \bar{l}\gamma$ production
             at three different LHC energies. The lower (upper) set is for
 MSTW 2008 (CTEQ6) parton distribution functions.
        }
\label{fig:ktot}
\end{figure}
%%%%%%%%%%%%%%%

We now turn to the
  dependence on the choice of the parton distributions. As figure
  \ref{fig:ktot} shows, the
variation of $K$-factor is about
$20\%-30\%$ for both the PDFs, CTEQ6 and MSTW 2008. The major
difference in $K$-factor between the two PDFs (specially at low center
of mass energy) is due to the different parametrization of their
parton distribution function 
(owing to their use of different data sets to extract the PDFs).
As can be expected, the
  difference is minor for low values of $M_* / \sqrt{S}$ (where
  experimental data abounds and the understanding is better), and
  increases with the ratio. This difference is irreducible at present
  and can be reduced only on inclusion of either more data (and,
  hence, more refined PDFs), or the calculation of still higher order
  effects.

%-------------
\begin{figure}[htb]
%\vspace{0.5cm}
\centerline{%\hspace*{3.0em}
\epsfxsize=18cm\epsfysize=14cm
                     \epsfbox{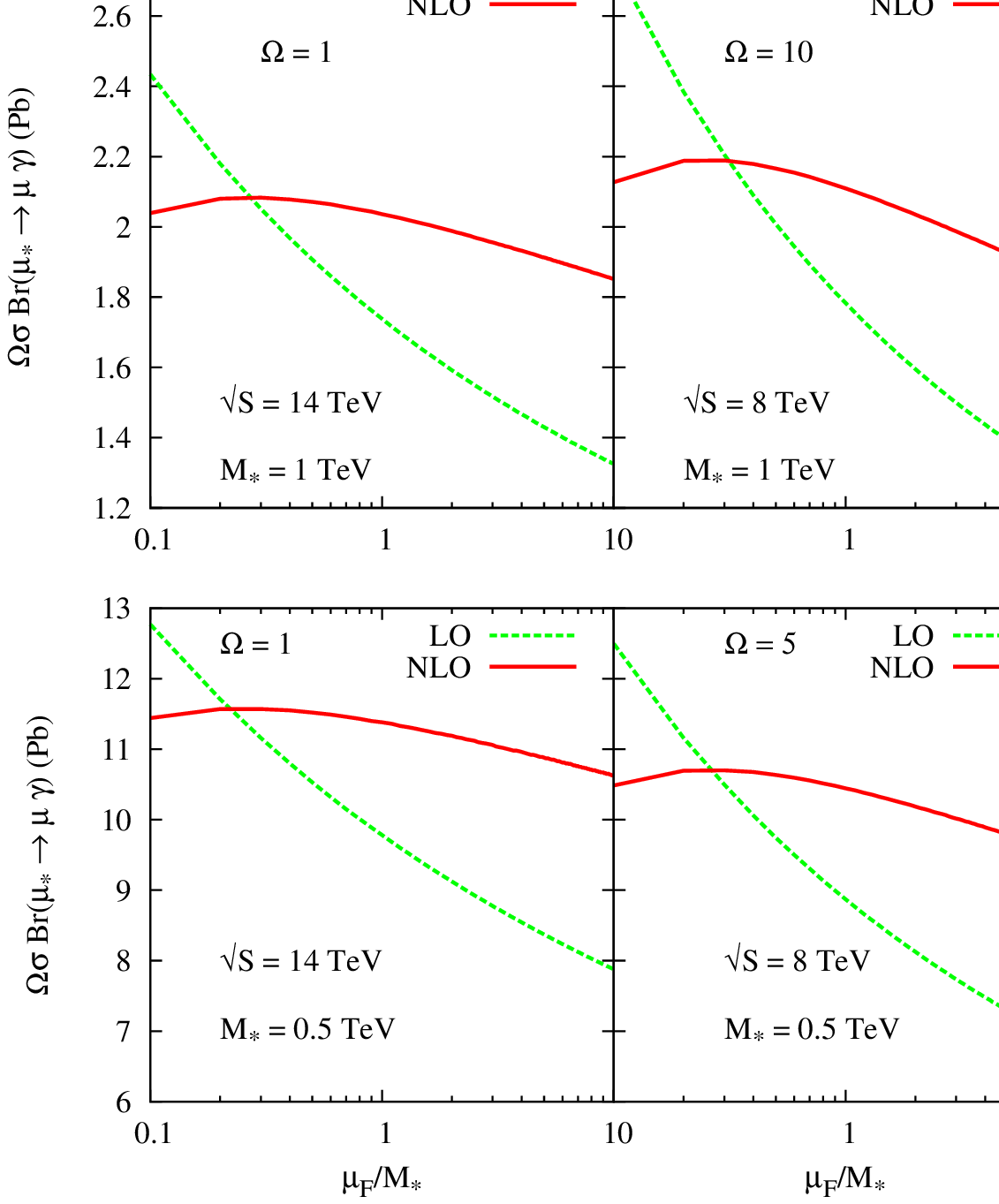}
}
\vspace{-1cm}
\caption{\em Variation of total cross section with respect to the factorization scale $\mu_F$ using CTEQ6 PDFs for $\Lambda = 2$ TeV. Here $\Omega$ is just a constant scale factor 
introduced to put all the graphs on same frame of respective scale. 
        }
\label{fig:scal_var}
\end{figure}
%------------

         In our above discussions, we have considered the simplest case 
$\mu_F^2=\mu^2_R=Q^2$ where the
cross section depends only on physical scales like 
the c.o.m. energy ($\sqrt{S}$)
and the masses of final state particles ($M_*$).
Now we turn on another scale called factorization scale $\mu_F^2 (= \mu_R^2$
the renormalization scale) and we have shown the factorization scale 
dependent of our NLO result in figure \ref{fig:scal_var}. From this figure,
it is clear that the scale dependence reduces greatly at NLO cross section 
compare
to LO cross section. This signifies the necessity of NLO QCD corrections.

\section{Conclusions}
\label{conclusion}

To conclude, we have systematically performed 
the next-to-leading order QCD corrections for the $V\pm A$ type
contact interactions as given in eqn.({\ref{lagrangianC}}). 
As opposed to naive expectations, we have showed that 
the QCD corrections are meaningful and reliable 
even in such a non-renormalizable theory.

We have analyzed the variation of cross section with respect to
the excited lepton mass (and, hence, the invariant mass of one SM lepton
and a SM gauge boson) at the LHC.
The enhancement of 
the NLO cross section over the LO cross section is 
found to be quite significant. To quantify the enhancement, we present 
the corresponding $K$-factors  
in a 
form suitable for experimental 
analyses.
A quick estimate shows that the inclusion of these corrections
changes the mass exclusion limits by about 60 GeVs.

As is well known, the 
cross section calculated at 
the leading order in perturbation theory suffers 
scale uncertainty 
on account of the arbitrariness in the
choice of factorization 
and renormalization scales.
These 
uncertainties are due to the absence of higher order contributions 
in the calculations.
On inclusion of each higher order,
these scale uncertainties
reduce gradually and the predictions 
are expected to become more reliable.
This is explicitly borne out by our calculations, which demonstrate 
that the scale dependence of the NLO 
result is greatly reduced in comparison to that 
for the LO case.

\section*{Acknowledgments} 
Author would like to thank Debajyoti Choudhury for useful discussions and comments. 
Author also wishes to acknowledge Satyaki Bhattacharya for useful discussions.

%%%%%%%%%%%%%%%%%%%%%%%%%%%%%%%%%%%%%%%%%%%%%%%%%%%%%%%%%%%
%\end{document}


\begin{thebibliography}{99}

\bibitem{susy}
%\cite{Nilles:1983ge}
% \bibitem{Nilles:1983ge}
  H.~P.~Nilles,
  %``Supersymmetry, Supergravity And Particle Physics,''
  Phys.\ Rept.\  {\bf 110} (1984) 1;\\
  %%CITATION = PRPLC,110,1;%%
%\cite{Haber:1984rc}
%\bibitem{Haber:1984rc}
  H.~E.~Haber and G.~L.~Kane,
  %``The Search For Supersymmetry: Probing Physics Beyond The Standard Model,''
  Phys.\ Rept.\  {\bf 117} (1985) 75;\\
  {\em Perspectives in Supersymmetry}, ed. G.L. Kane,
World Scientific (1998); \\
  {\em Theory and Phenomenology of Sparticles}:
M. Drees, R.M. Godbole and P. Roy, World Scientific (2005).
  %%CITATION = PRPLC,117,75;%%


%\cite{Pati:1974yy}
 \bibitem{Pati:1974yy}
  J.~C.~Pati and A.~Salam,
  %``Lepton Number As The Fourth Color,''
  Phys.\ Rev.\ D {\bf 10} (1974) 275.
  %%CITATION = PHRVA,D10,275;%%

\bibitem{GUTS}
%\cite{Georgi:1974sy}
% \bibitem{Georgi:1974sy}
  H.~Georgi and S.~L.~Glashow,
  %``Unity Of All Elementary Particle Forces,''
  Phys.\ Rev.\ Lett.\  {\bf 32} (1974) 438 ;\\
  %%CITATION = PRLTA,32,438;%%
%\cite{Langacker:1980js}
% \bibitem{Langacker:1980js}
  P.~Langacker,
  %``Grand Unified Theories And Proton Decay,''
  Phys.\ Rept.\  {\bf 72} (1981) 185.
  %%CITATION = PRPLC,72,185;%%



\bibitem{llmodel}
E. Eichten, K.D. Lane and M.E. Peskin, Phys. Rev. Lett. {\bf 50} (1983) 811 ;\\
E. Eichten, I. Hinchliffe, K.D. Lane and C. Quigg, Rev. Mod. Phys.
{\bf 56} (1984) 579. 

\bibitem{comp_mod}
Jogesh C. Pati, Abdus Salam and J.A. Strathdee Phys.Lett. {\bf B59} (1975) 265;\\
H. Fritzsch and G. Mandelbaum, Phys.Lett. {\bf B102} (1981) 319;\\
W. Buchmuller, R.D. Peccei and T. Yanagida, Phys.Lett. {\bf B124} (1983) 67;
Nucl.Phys.{\bf B227} (1983) 503; Nucl.Phys.{\bf B237} (1984) 53;\\
 U. Baur and H. Fritzsch, Phys.Lett. {\bf B134} (1984) 105;\\
Xiaoyuan Li and R.E. Marshak, Nucl.Phys.{\bf B268} (1986) 383;\\
 I. Bars, J.F. Gunion and M. Kwan Nucl.Phys.{\bf B269} (1986) 421;\\
 G. Domokos and S. Kovesi-Domokos, Phys.Lett.{\bf B266} (1991) 87;\\
Jonathan L. Rosner and Davison E. Soper Phys.Rev.{\bf D45} (1992) 3206;\\
Markus A. Luty and Rabindra N. Mohapatra, Phys.Lett.{\bf B396} (1997) 161 [hep-ph/9611343];\\
K. Hagiwara, K. Hikasa and M. Tanabashi, Phys.Rev.{\bf D66} (2002) 010001; Phys.Lett.{\bf B592} (2004) 1.

\bibitem{Additional_comp}
For a review and additional references, see R.R. Volkas and G.C. Joshi, Phys. Rep.
{\bf 159} (1988) 303.


\bibitem{preon}
H. Harari and N. Seiberg, Phys.Lett. {\bf B98} (1981) 269;\\
M.E.  Peskin, {\it in proceedings of the 1981 International Symposium on Lepton and Photon Interaction at High Energy,}
W.Pfeil, ed., p880 (Bonn, 1981);\\
L. Lyons, Oxford University Publication 52/82 (June 1982);\\
%
%\bibitem{preon_binding}
G. 't Hooft, in Recent Developements in Gauge Theories;\\
G. 't Hooft {\it et al.}, ads. (Plenum Press, New York,1980).

\bibitem{LagrangianCI}
U. Baur, M. Spira, and P. Zerwas, Phys. Rev. D{\bf 42} (1990) 815;\\
J.Kuhn and P. Zerwas, Phys.Lett. {\bf B147} (1984) 189.

\bibitem{LagrangianGM}
F. Boudjema, A. Djouadi, and J. Kneur, Z. Phys. C{\bf 57} (1993) 425;\\
K. Hagiwara, D. Zeppenfeld and S. Komamiya, Z. Phys. C{\bf 29}, 115 (1985);\\
N. Cabibbo, L. Maiani and Y. Srivastava, Phys. Lett. B{\bf 139}, 459 (1984).

%\bibitem{leptoquarks}
%  W.~Buchmuller and D.~Wyler,
%  Phys.\ Lett.\ B {\bf 177} (1986) 377;\\
%  W.~Buchmuller, R.~Ruckl and D.~Wyler,
%  Phys.\ Lett.\ B {\bf 191} (1987) 442;
%  [Erratum-ibid.\ B {\bf 448} (1999) 320 ];\\
%  J.~L.~Hewett and T.~G.~Rizzo,
%  Phys.\ Rev.\ D {\bf 56} (1997) 5709
%  [hep-ph/9703337].

%\bibitem{fayet} 
%  P.~Fayet,
%  Phys.\ Lett.\ B {\bf 69}  (1977) 489;\\
%  G.~R.~Farrar and P.~Fayet,
%  Phys.\ Lett.\ B {\bf 76} (1978) 575.
%
%\bibitem{Hasenfratz:1987tk}
%  P.~Hasenfratz and J.~Nager,
%  Z.\ Phys.\ C {\bf 37} (1988) 477.
%
\bibitem{Delphi}
ALEPH Collaboration, Phys. Lett. B{\bf 385} (1996) 445; \\
OPAL Collab., Eur. Phys. J. C{\bf 14} (2000) 73; \\
L3 Collab., Phys. Lett. B{\bf 568} (2003) 23; \\
DELPHI Collaboration, Eur.Phys. J. {\bf C 8} (1999) 41; Eur. Phys. J. C{\bf 46} (2006) 277.

\bibitem{HERA2}
H1 Collaboration, Phys.Lett.{\bf B678} (2009)335;
Phys.Lett.{\bf B666} (2008) 131; Eur. Phys. J. {\bf C 17} (2000) 567; \\
ZEUS Collaboration S. Chekanov {\it et al.}, Phys. Lett. B {\bf 549} (2002) 32.

\bibitem{cdfprl}
CDF Collaboration, Phys. Rev. Lett. {\bf 94} (2005) 101802; Phys. Rev. Lett. {\bf 97} (2006) 191802; \\
D0 Collaboration,Phys. Rev. D{\bf 73} (2006) 111102; Phys. Rev. D{\bf 77} (2008) 091102.

\bibitem{CMS_CI} CMS Collaboration, Phys.Lett. {\bf B704} (2011) 143. 

\bibitem{ATLAS_CI} ATLAS Collaboration, Phys.Rev. {\bf D85} (2012) 072003.

\bibitem{TauBR} J.L. Diaz and O.A. Sampayo, Phys.Rev. {\bf D49} (1994) R2149.

\bibitem{TauBR_bounds} J.I. Aranda, R. Martinez and O.A. Sampayo, Phys.Rev. {\bf D62} (2000) 013010.

\bibitem{Drell_Yan} 
S.D. Drell and T.M. Yan, Phys. Rev. Lett. {\bf 25} (1970) 316;\\
J.H. Christenson {\it et al.}, {\it ibid.} {\bf 25} (1970) 1523;\\
L.M. Lederman and B.G. Pope, {\it ibid.} {\bf 27} (1971) 765.

\bibitem{NLL} 
R. Hamberg, W. L. van Neerven and T. Matsuura, Nucl. Phys. B{\bf 359} (1991) 343 .

\bibitem{Martin}
P.J. Sutton {\em et al.}, Phys. Rev. {D45} (1992) 2349;\\
 A.D. Martin {\em et al.}, Phys. Lett. {B 354} (1995) 155 
[hep-ph/9502336].

\bibitem{JETRAD} W.T. Giele, E.W.N. Glover and D.A. Kosower, Nucl. Phys. {\bf B403}
(1993) 633 [hep-ph/9302225].

\bibitem{HERWIG} G. Corcella, I.G. Knowles, G. Marchesini, S. Moretti, K. Odagiri, P. Richardson, M.H. Seymour and B.R. Webber, JHEP{\bf 0101} (2001) 010.

%\bibitem{NLL} 
%R. Hamberg, W. L. van Neerven and T. Matsuura, Nucl. Phys. B{\bf 359} (1991) 343 .

%\bibitem{Martin}
%P.J. Sutton {\em et al.}, Phys. Rev. {D45} (1992) 2349;\\
% A.D. Martin {\em et al.}, Phys. Lett. {B 354} (1995) 155 
%[hep-ph/9502336].

%\bibitem{CMS_CI} CMS Collaboration, Phys.Lett. {\bf B704} (2011) 143. 

%\bibitem{ATLAS_CI} ATLAS Collaboration, Phys.Rev. {\bf D85} (2012) 072003.

\bibitem{SM_CI} D. Choudhury, S. Majhi and V. Ravindran, JHEP {\bf 0601} (2006) 027.

\bibitem{ravi_gravi1} 
 P. Mathews, V. Ravindran, K. Sridhar and W.L. van Neerven,
Nucl. Phys. {\bf B713} (2005) 333 [hep-ph/0411018].

\bibitem{CI_3BDK} S. Majhi {\it et al.} work in progress.

\bibitem{ravi_gravi} 
G. Altarelli, R.K.Ellis and G. Martinelli,
Nucl. Phys. {\bf B157} (1979) 461;\\
B.Humpert and W.L. van Neerven, Phys. Lett. {\bf B84} (1979) 327;
[Errat. {\bf B85} (1979) 471]; ibid. {\bf B89} (1979) 69; Nucl. Phys. {\bf B
184} (1981) 225;\\
J.Kubar, M. le Bellac, J.L.Meunier and G. Plaut, Nucl. Phys. {\bf B175}
(1980) 251;\\
P. Aurenche and P. Chiapetta, Z.Phys. {\bf C34} (1987) 201;\\
P.J.Sutton, A.D.Martin, R.G. Roberts W.J.Stirling, Phys. Rev. {\bf D45}
(1992) 2349;\\
P.J. Rijken and W.L. van Neerven, Phys. Rev. {\bf D51} (1995) 44
[hep-ph/9408366].

%\bibitem{ravi_gravi1} 
% P. Mathews, V. Ravindran, K. Sridhar and W.L. van Neerven,
%Nucl. Phys. {\bf B713} (2005) 333 [hep-ph/0411018].

%\bibitem{BN_theory}
%F. Block and A. Nordsieck, Phys. Rev. {\bf 52} (1937) 54;\\
%D.R. Yannie, S.C. Frautschi and H. Suura, Ann. Phys. (N.Y.) {\bf 13} (1961) 379.

%\bibitem{mass_fac}
%T. Kinoshita, J. Math. Phys. {\bf 3} (1962) 650;\\
%T.D. Lee and M. Nauenberg, Phys. Rev. {\bf 133} (1964) B1549;\\
%N. Nakanishi Prog. Theor. Phys. {\bf 19} (1958) 159.

\bibitem{CTEQ6}
J. Pumplin, D.R. Stump, J. Huston, H.L. Lai, Pavel M. Nadolsky and W.K. Tung,
JHEP {\bf 0207} (2002) 012.

\bibitem{PRD65} O.J.P. Eboli, S.M. Lietti and P. Mathews, Phys.Rev. {\bf D65} (2002) 075003.

\bibitem{PRD81} S. C. Inan, Phys.Rev. {\bf D81} (2010) 115002. 

\bibitem{MSTW}
A.D. Martin, W.J. Stirling, R.S. Thorne and G. Watt, Eur. Phys. J. {\bf C63} 
(2009) 189–285.


\end{thebibliography}
\end{document}